%% file: lgi.tex
\documentclass[twocolumn,pra,aps,showpacs,superscriptaddress,nofootinbib]{revtex4}

\usepackage{amsmath}
\usepackage{graphicx}
\usepackage{hyperref}
\usepackage{xspace}
\usepackage{bm}
\usepackage{float}
\usepackage{afterpage}
\usepackage{placeins}

\makeatletter
\gdef\@fpheader{}
\g@addto@macro\bfseries{\boldmath}
\makeatother

\hypersetup{
    bookmarks=true,         
    unicode=false,          
    pdftoolbar=true,        
    pdfmenubar=true,        
    pdffitwindow=false,     
    pdfstartview={FitH},    
    pdfnewwindow=true,      
    colorlinks=true,        
    linkcolor=red,          
    citecolor=cyan,         
    filecolor=magenta,      
    urlcolor=blue,         
    linktocpage=true
}

\input{newcommands}

\begin{document}

\title{Leggett-Garg Inequalities for Squeezed States}

\author{J\'er\^ome Martin} 
\email{jmartin@iap.fr}
\affiliation{Institut d'Astrophysique de Paris, UMR 7095-CNRS,
Universit\'e Pierre et Marie Curie, 98 bis boulevard Arago, 75014
Paris, France}

\author{Vincent Vennin} 
\email{vincent.vennin@port.ac.uk}
\affiliation{Institute of Cosmology and Gravitation, University of
  Portsmouth, Dennis Sciama Building, Burnaby Road, Portsmouth, PO1
  3FX, United Kingdom}

\date{\today}

\begin{abstract}
  Temporal Bell inequalities, or Leggett-Garg Inequalities (LGI), are
  studied for continuous-variable systems placed in a squeezed
  state. The importance of those systems lies in their broad
  applicability, which allows the description of many different
  physical settings in various branches of physics, ranging from
  cosmology to condensed matter physics and from optics to quantum
  information theory. LGI violations are explored and systematically
  mapped in squeezing parameter space. Configurations for which LGI
  violation occurs are found, but it is shown that no violation can be
  obtained if all squeezing angles vanish, contrary to what happens
  for the spatial Bell inequalities. We also assess the effect of
  decoherence on the detectability of such violations. Our study opens
  up the possibility of new experimental designs for the observation
  of LGI violation.
\end{abstract}

\maketitle

\section{Introduction}
One of the most interesting aspects of quantum mechanics is the
possibility of having entangled
states~\cite{Einstein:1935rr}. Surprising properties of these states
are revealed by Bell's inequalities~\cite{Bell:1964kc,Clauser:1969ny},
which highlight the non-classical correlations that can exist between
spatially-separated
sub-systems~\cite{Aspect:1982fx,Aspect:1981nv}. However, quantum
  mechanics may also imply the presence of non-standard correlations
when a single system is measured at two different times. These
correlations can be studied by deriving another class of inequalities,
known as temporal Bell inequalities or Leggett-Garg inequalities
(LGI)~\cite{PhysRevLett.54.857}. A violation of these inequalities can
be caused by the lack of a realistic description of the system
(macroscopic realism) or by the impossibility to measure its
properties without disturbing it (non-invasive
measurability). Therefore, LGI violations have deep and far-reaching
implications for quantum mechanics and physics in
general~\cite{2014RPPh...77a6001E}.

On the experimental side, violations of LGI have now been observed in
different systems, for a review see
Ref.~\cite{2014RPPh...77a6001E}. These systems all share the property
of being effectively describable as a qubit. Originally, it was
proposed to use a rf SQUID to perform the
test~\cite{PhysRevLett.54.857}. The rf SQUID is a magnetic flux box
made of a Josephson junction inserted into a superconducting ring and
controlled by an external flux.  If the external flux is tuned to half
the quantum flux, then the potential of the system acquires a
double-well shape, the lowest energy states in each well effectively
decouple, and the rf SQUID can be viewed as a qubit. The two states
correspond to clockwise and anticlockwise super-current states. In
practice, the first LGI violation was in fact observed in a
transmon~\cite{2010NatPh...6..442P}, a device similar to a Cooper-pair
box (a Josephson junction driven by an applied voltage) but operated
in a different regime~\cite{PhysRevA.76.042319}.
In that case, the qubit is represented by the $\vert 0\rangle $ and
$\vert 1\rangle $ charge states. Since this experiment, many
  other protocols have been considered, in particular the recent
  proposals studied
  in \Refs{PhysRevLett.111.090506,2016arXiv160103728K}. Let us
  note that LGI violations have also been observed
  for other systems~\cite{2014RPPh...77a6001E} equivalent to a qubit but not
  necessarily based on superconducting circuits.

The non-invasive measurability principle, which has been the subject
of many debates, has also played an important role in the design of
the different experimental systems. In the original
proposal~\cite{PhysRevLett.54.857}, the measurements were assumed to be
performed in the usual, projective, fashion. However, in the first
experimental realization of \Refc{2010NatPh...6..442P},
continuous weak measurements were used, following
\Refc{2006PhRvL..96t0404R}, which required one to adapt the original
form of the LGI to the situation at hand. The subsequent experimental
setups then considered various variations of the measurement
protocol.

In this paper we propose and study a new and generic way to design
physical situations in which the LGI are violated. Concretely, our
approach applies to any continuous-variable system placed in a quantum
squeezed state~\cite{Caves:1985zz,Schumaker:1985zz}. These states are
entangled states and arise in a large variety of physical
situations. The reason is that any Hamiltonian that is bilinear in the
creation and annihilation operators is likely to produce squeezed
states. One finds them in experiments with light fields using lasers
and non-linear optics (parametric down-conversion, four-wave
mixing)~\cite{PhysRevLett.57.2520,Eberle:2010zz} or experiments
probing the motion of an ion in a trap or the properties of phonons in
a crystal~\cite{2011PhLA..375.4141M}. Squeezed states are also
unavoidable when a quantum field interacts with a classical source as
is the case in the Schwinger~\cite{Schwinger:1951nm},
Unruh~\cite{Unruh:1976db} and Hawking~\cite{Hawking:1974sw}
effects. Moreover, according to the theory of cosmic
inflation~\cite{Starobinsky:1980te,Guth:1980zm,Linde:1981mu}, recently
confirmed by the Planck satellite
data~\cite{Ade:2015xua,Ade:2015lrj,Martin:2013tda,Martin:2013nzq,Martin:2014nya,Martin:2015dha},
the quantum state of the cosmological fluctuations responsible for the
Cosmic Microwave Background Radiation (CMBR) anisotropy and the large
scale structures observed in our universe, is a (two-mode) squeezed
state~\cite{Grishchuk:1990bj,Martin:2007bw,Martin:2012pea,Martin:2015qta}. Let
us add that squeezed states are also very useful for interferometric
measurements (which, for instance, are used for the detection of
gravitational waves~\cite{TheLIGOScientific:2013nha}) or to improve
the precision of atomic clocks~\cite{PhysRevLett.104.250801}. They
also play a crucial role in quantum information
processing~\cite{RevModPhys.77.513}. Finally, it is interesting to
note that entangled states were introduced for the first time in
their squeezed state realization since the Einstein-Podolski-Rosen
(EPR) state~\cite{Einstein:1935rr} is nothing but a squeezed state
with infinite squeezing.

Let us also stress that a fundamental difference with qubits is
  that squeezed states describe continuous variable systems (see also
  Ref.~\cite{2015PhRvL.115t0403B}). Besides opening up possibilities
  of new experimental LGI violations, this may also provide a way
  to test the quantumness of primordial cosmological fluctuations~\cite{Martin:2015qta}, the possibility of which is still being debated.

This paper is organized as follows. In
\Sec{sec:spinop}, we introduce spin operators for
continuous-variable systems and calculate their two-point correlation
functions. In \Sec{sec:lgi}, we then show that LGI can be
violated and we map the LGI violations in the space of the squeezing
parameters and squeezing angles. Finally, we present our conclusions 
in \Sec{sec:conclusion}. The appendix~\ref{sec:2pt} contains 
technical details needed to derive the results discussed in the 
main text.

\section{Spin operators for squeezed states}
\label{sec:spinop}
We consider a quantum system that possesses continuous degrees of
freedom denoted in the following by $Q$; $Q$ could, for instance, be
the position of a particle or the Fourier amplitude of a field at a
given wave number. In order to test the LGI, one needs to define a
dichotomic quantity and, for this reason, we introduce the following
operator
\begin{align}
\label{eq:defsz}
\hat{S}_z(\ell)
& =\sum_{n=-\infty}^{n=+\infty}(-1)^n\int _{n\ell}
^{(n+1)\ell}{\rm d}Q \vert Q\rangle 
\langle Q\vert \, ,
\end{align}
where $\ell$ is a parameter that can be freely chosen by the
  observer and that describes the coarseness of the measurement (the larger $\ell$, the coarser the measurement). 
In the limit where $\ell\rightarrow \infty$, for instance, $\hat{S}_z(\ell)$ is simply the sign of $Q$. In general, \Eq{eq:defsz} 
defines a spin variable because the eigenvalues of this operator are
$\pm 1$ (for an alternative way of defining dichotomic variables
  see Ref.~\cite{2016PhRvA..94a2105L}). It is similar to the
$z$-component of a fictitious spin. As explained in
Ref.~\cite{2004PhRvA..70b2102L}, one could define two other operators,
$\hat{S}_x(\ell)$ and $\hat{S}_y(\ell)$ such that $\hat{S}_x(\ell)$,
$\hat{S}_y(\ell)$ and $\hat{S}_z(\ell)$ obey the standard $SU(2)$
commutation relations. Here, we will only need $\hat{S}_z(\ell)$. It
is not the only way to define a ``spin'' from a continuous-variable
system and we could also consider the
operator~\cite{PhysRevLett.82.2009,PhysRevLett.88.040406}
$\sum_{n=0}^{+\infty}\left(\vert 2n+1\rangle \langle 2n+1\vert -\vert
  2n \rangle \langle 2n\vert\right)=(-1)^{\hat{N}+1}$ where $\vert
n\rangle$ are the Fock states and $\hat{N}$ is the number
operator. However, in this case, it is not obvious how to design an
experimental protocol to perform a measurement of $(-1)^{\hat{N}+1}$
while measuring the quantity~(\ref{eq:defsz}) is straightforward since
only ``position'' measurements are needed. In this sense, the
choice~(\ref{eq:defsz}) of $\hat{S}_z(\ell)$ appears to be essentially
unique. Note that Bell inequalities formed out of the triplet
$\hat{S}_x(\ell)$, $\hat{S}_y(\ell)$ and $\hat{S}_z(\ell)$ can be
violated if the system is placed in a squeezed state as recently shown
in Refs.~\cite{2004PhRvA..70b2102L,Martin:2016tbd}. However, this
involves the measurements of, at least, two spin operators and
measuring $\hat{S}_x(\ell)$ and/or $\hat{S}_y(\ell)$ requires the
measurement of the conjugate momentum of $\hat{Q}$, which may be difficult
(\eg in inflationary cosmology, this appears to be essentially
impossible~\cite{Martin:2015qta}). The advantage of the LGI is that
only one operator is necessary, the price to pay being of course that
one has to measure it at three different times.

Then, we assume that the system is placed in the quantum state
$\hat{U}\vert 0\rangle $, a (one-mode) squeezed state, where the
evolution operator $\hat{U}$ can be written as
$\hat{U}(t)=e^{\hat{B}(t)}$ with $ \hat{B}(t)\equiv
r(t)e^{-2i\varphi(t)}\hat{a}^2/2 -r(t)e^{2i\varphi(t)}
\left(\hat{a}^{\dagger}\right)^2/2$, where $\hat{a}$ and
$\hat{a}^{\dagger}$ are, respectively, the annihilation and creation operators
satisfying $[\hat{a},\hat{a}^{\dagger}]=1$. The quantities $r(t)$ and
$\varphi(t)$ are respectively the squeezing parameter and angle and
are typically time-dependent quantities.
\begin{figure}[t]
\begin{center}
\includegraphics[width=0.35\textwidth,clip=true]{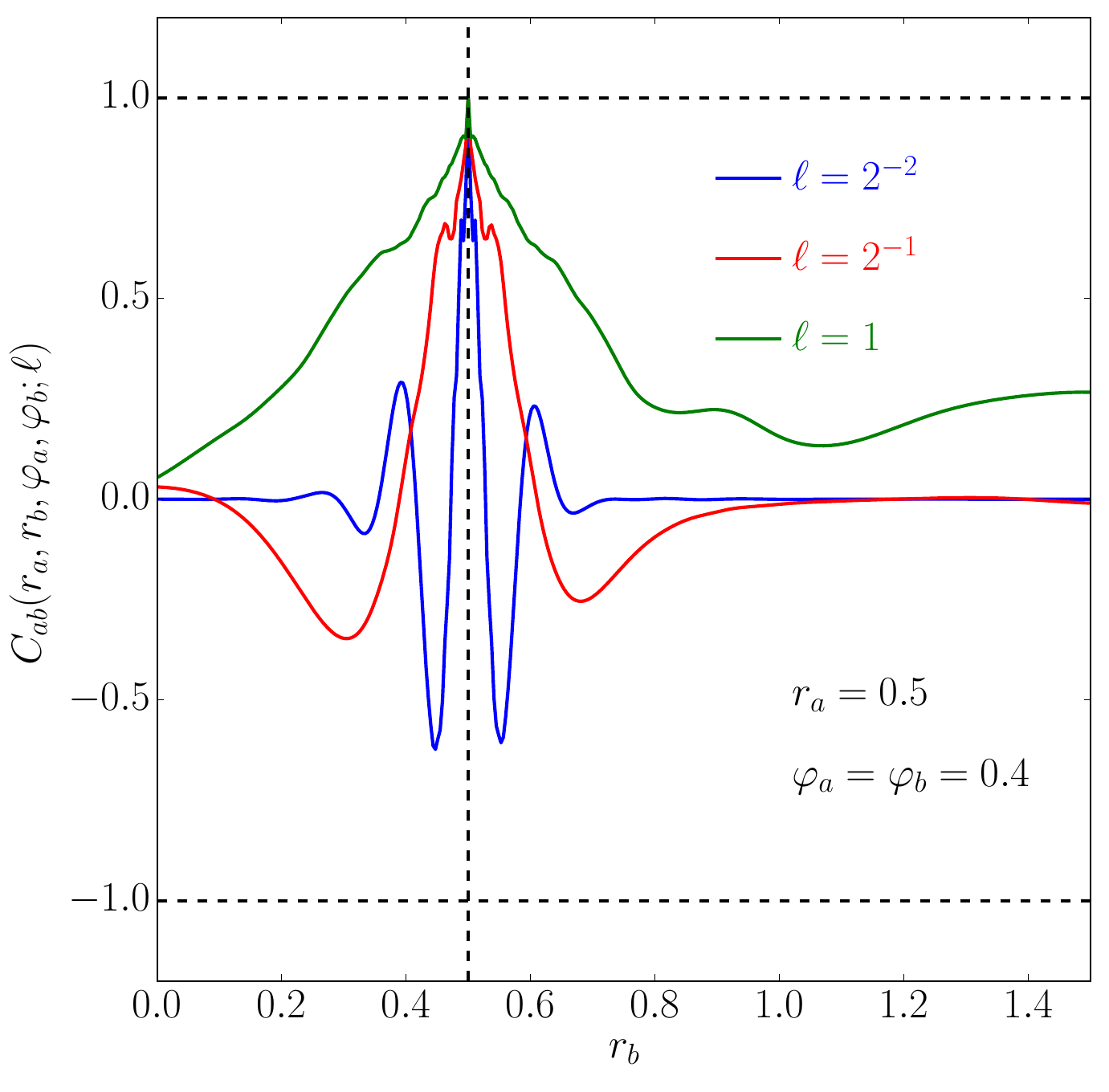}
\includegraphics[width=0.35\textwidth,clip=true]{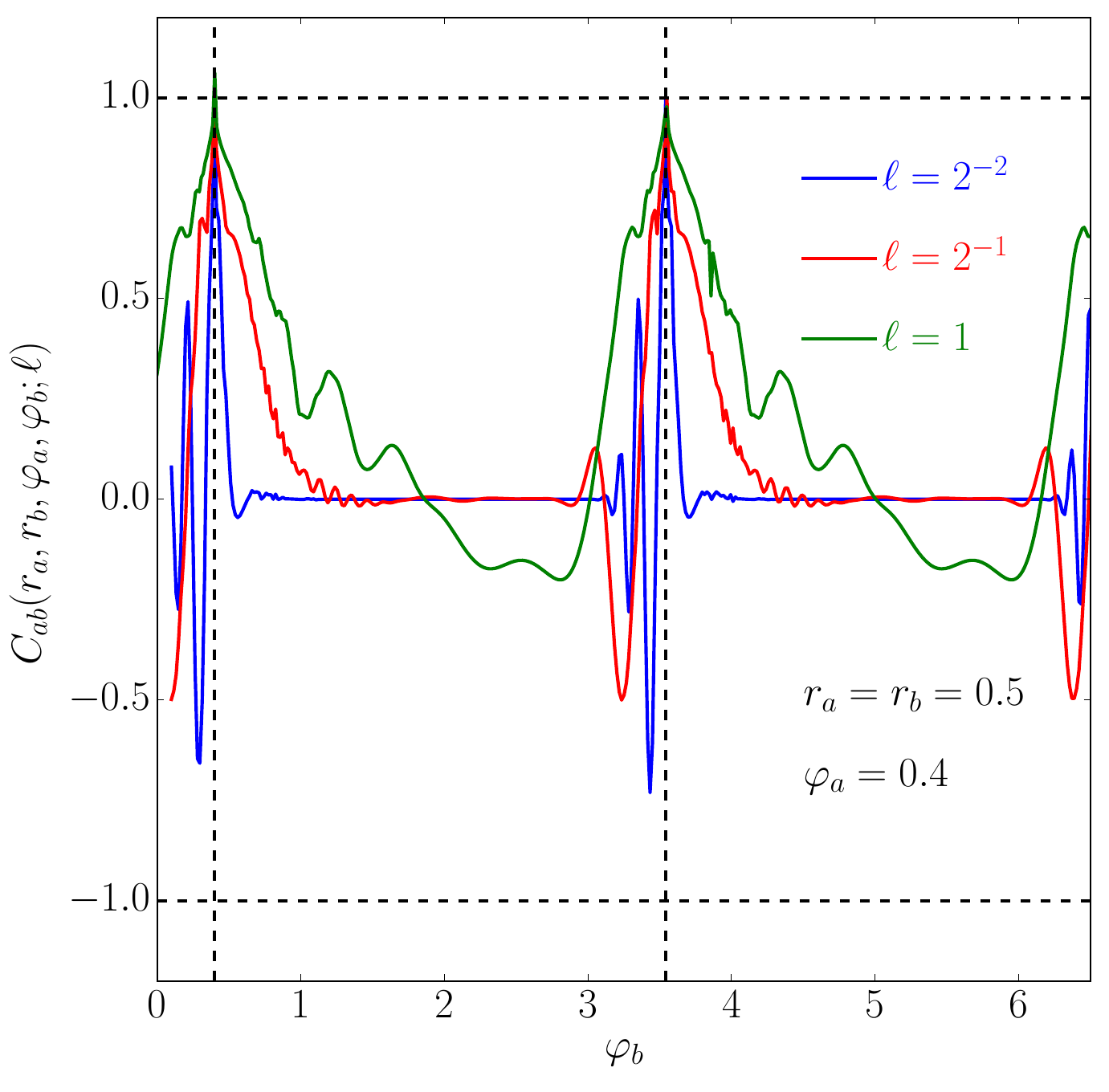}
\caption{Two-point function of the spin
  operator, the system being placed in a squeezed
  state with squeezing parameter and angle $(r_a,\varphi_a)$ at time
  $t_a$ and $(r_b,\varphi_b)$ at time $t_b$, as a function of $r_b$
  (top panel) and $\varphi_b$ (bottom panel), for a few values of
  $\ell$.}
\label{fig:correlation_r}
\end{center}
\end{figure}
\begin{figure}[t]
\begin{center}
\includegraphics[width=0.343\textwidth,clip=true]{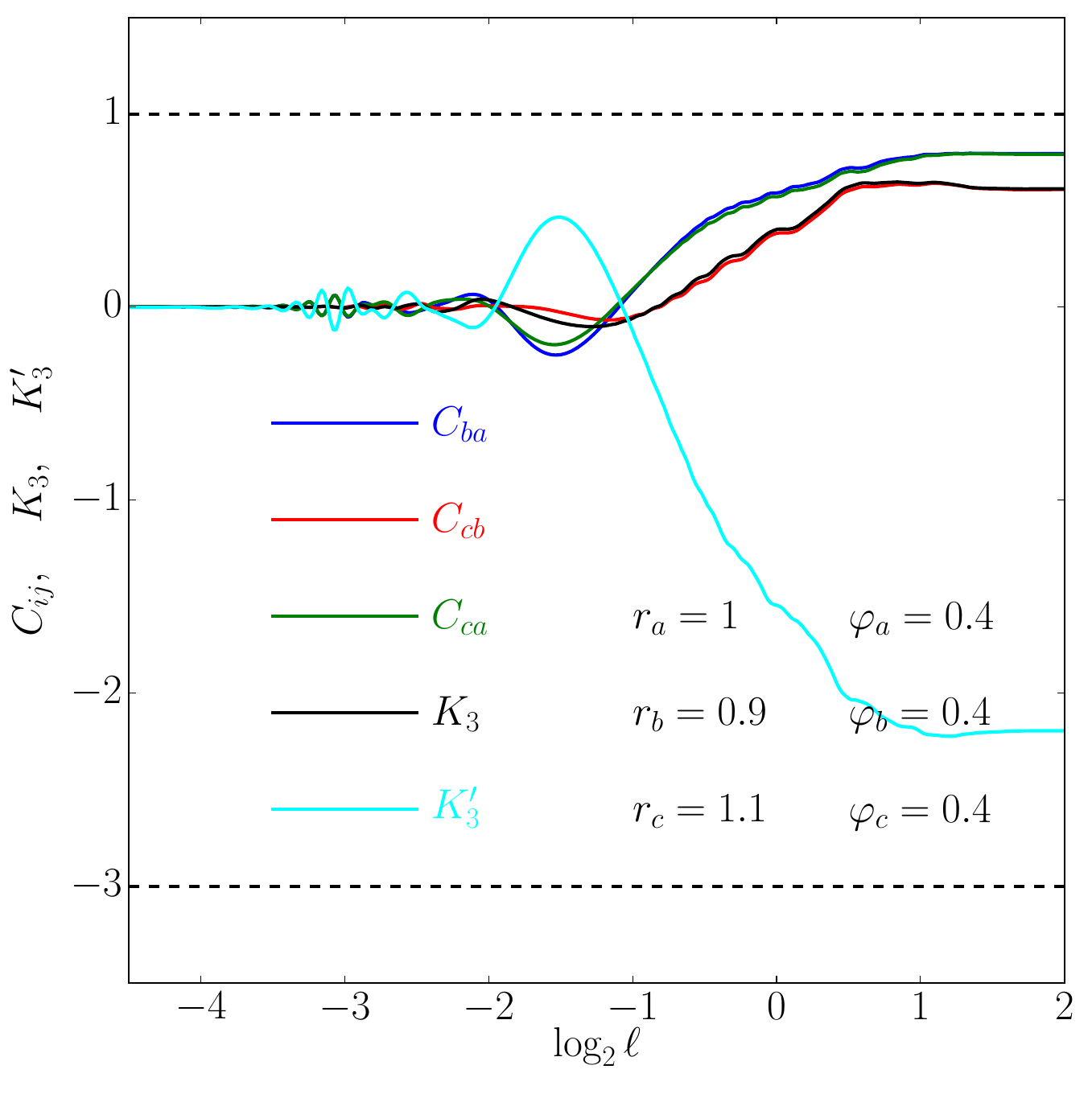}
\includegraphics[width=0.343\textwidth,clip=true]{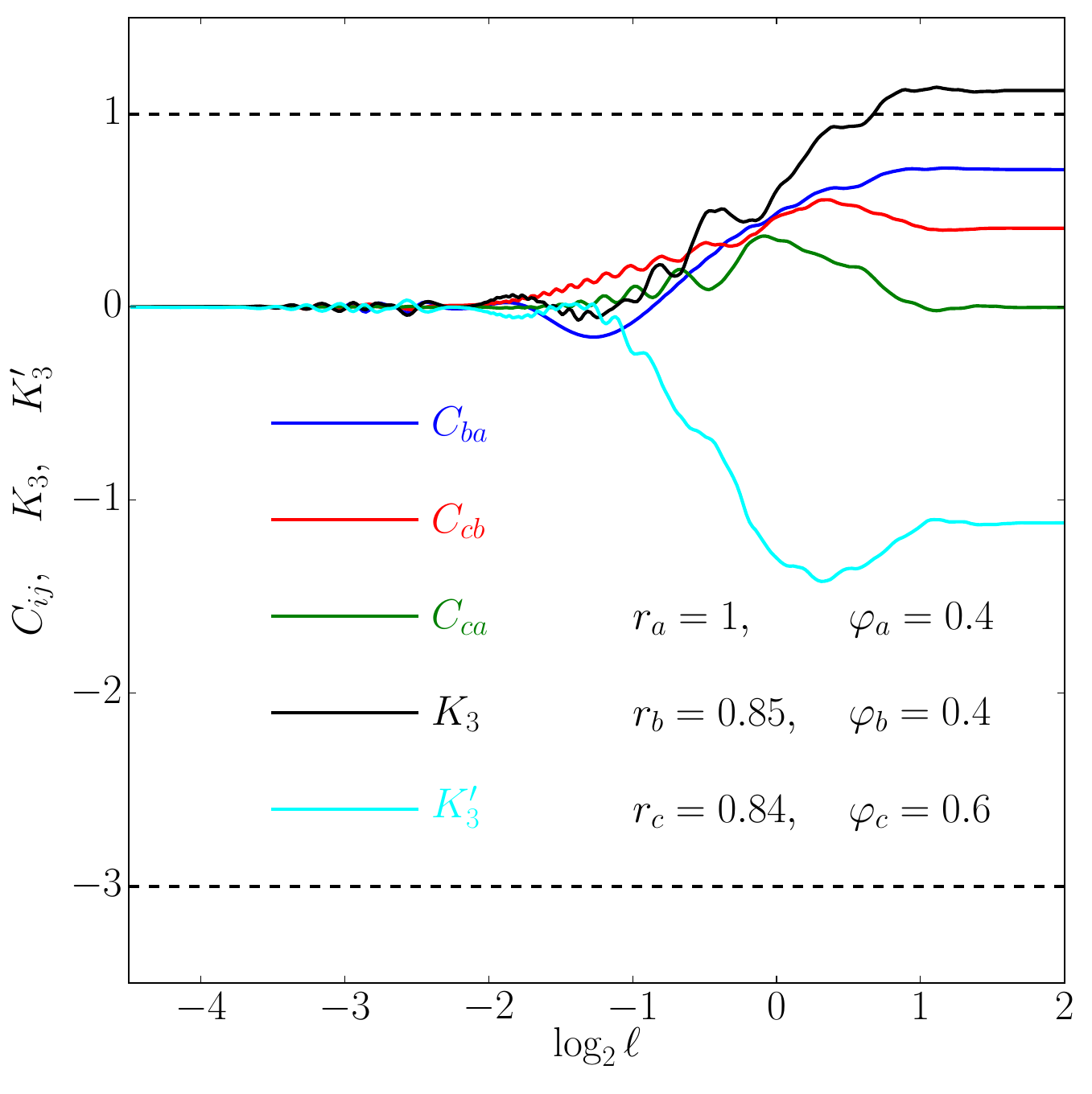}
\caption{Two-point function~(\ref{eq:defcorrel}) of the spin
  operator~(\ref{eq:defsz}) and Leggett-Garg strings~(\ref{eq:K3:def})
  for the squeezed state measured at times $t_a$, $t_b$ and $t_c$, as
  a function of $\ell$. LGI violations correspond to $K_3$ or
  $K_3^\prime$ being either smaller than $-3$ or greater than $1$. For
  the parameters used in the bottom panel for instance, violation
  $K_3>1$ occurs at large $\ell$. }
\label{fig:k3}
\end{center}
\end{figure}

In the following, we will be interested in the calculation of the
two-point correlation function of the spin operator~(\ref{eq:defsz})
taken at different times, namely
\begin{equation}
\label{eq:defcorrel}
C_{ab}(t_a,t_b;\ell)=\frac12 \left \langle 0\left\vert 
\left\{\hat{S}_z(t_a;\ell),\hat{S}_z(t_b;\ell)\right\}
\right \vert  0\right \rangle ,
\end{equation}
where $\hat{S}_z(t,\ell)$ is written in the Heisenberg picture, namely
$\hat{S}_z(t,\ell)\equiv
U^{\dagger}(t)\hat{S}_z(\ell)\hat{U}(t)$. This gives rise to (see
the appendix for the following formulas)
\begin{eqnarray}
\label{eq:Cab:Gauss}
C_{ab}& = &\Rea\left[{\cal A}(a,b)
\sum_{n=-\infty}^{n=+\infty}\sum_{m=-\infty}^{m=+\infty}
(-1)^{n+m}\int _{n\ell}^{(n+1)\ell}
\right. \nonumber \\ &  & \hspace{-1em} \left.
\int _{m\ell}^{(m+1)\ell}{\rm d}\tilde{Q}{\rm d}\overline{Q}
e^{A(a,b)\tilde{Q}^2+A^*(b,a)\overline{Q}^2+B(a,b)\overline{Q}\tilde{Q}}\right]
\end{eqnarray}
with (see the \textit{erratum} section at the very end)
\begin{eqnarray} 
{\cal A}(a,b) &=& \frac{1}{\pi\sqrt{2}} \frac{\sin
    ^{-1/2}(\theta_a-\theta_b)}{\rho_a\rho_b \cosh r_a \cosh r_b}
  e^{i(\theta_a-\theta_b-\pi/2)/2},
  \\
  \label{eq:A:def}
  A(a,b) &=& \frac12 -\frac{\cos \theta_a}{\rho_a}
  +\frac{i}{\rho_a}\frac{\cos \theta_b}{\sin (\theta_a -\theta_b)}
  \nonumber \\ & &
  -i\frac{\sin \theta_a}{\rho_a} -\frac{i}{2\tan (\theta_a-\theta_b)},        \\
  B(a,b) &=& -\frac{i}{\rho_a\rho_b\sin(\theta_a-\theta_b)}
  \frac{1}{\cosh r_a \cosh r_b},
\label{eq:B:def}
\end{eqnarray}
where $\rho$ and $\theta$ are defined so that $1-\tanh(r) e^{2 i
  \varphi}\equiv\rho e^{i\theta}$. In general, these integrations must be
performed numerically.\footnote{A \textsc{fortran} code for computing the two-point correlation function of the spin operators, the 3-measurement Leggett-Garg strings and all quantities displayed in this paper can be found at \url{https://github.com/vennin/LeggetGargInequalities}.} However, if we are considering the case where the
squeezing angles all vanish, then the integrations can be done and the series
reduces to $C_{ab}=\sum_{n=-\infty}^{n=+\infty} (-1)^nC_n$ with 
\begin{align}
\label{eq:cn1}
C_n = &\frac12
(-1)^{{\rm E}(e^{r_a-r_b}n)} \bigl\{2{\rm erf}\left[e^{r_b} {\rm
    E}(e^{r_a-r_b}n)\ell +e^{r_b}\ell\right] \nonumber \\
& -{\rm
  erf}\left[e^{r_a}(n+1)\ell)\right] -{\rm
  erf}\left(e^{r_a}n\ell\right)\bigr\},
\end{align}
if the condition $e^{r_b-r_a}[{\rm E}(e^{r_a-r_b}n)+1]<n+1$ is
satisfied, while
\begin{align}
\label{eq:cn2}
C_n = \frac12 (-1)^{{\rm E}(e^{r_a-r_b}n)}
\left\{{\rm erf}\left[e^{r_a}(n+1)\ell\right]
-{\rm erf}\left(e^{r_a}n\ell\right)\right\}
\end{align}
otherwise. In these expressions, ${\rm erf}$ is the error function and
${\rm E}(z)$ denotes the integer part of the number $z$. Another case
where a simple analytic expression can be derived is the limit
$\ell\rightarrow\infty$, where one obtains [see
  \Eq{eq:Cab:ellInfinite} in Appendix~\ref{sec:2pt}]
\begin{align}
C_{ab}= \Rea\left\lbrace
-\frac{4\mathcal{A}\, \mathrm{arctanh}\left[
B/\sqrt{B^2 - 4 A A^*} \right]}{\sqrt{B^2-4AA^*}}
\right\rbrace
\label{eq:Cab:ell_eq_infinity}
\end{align}
that is to say a plateau, independent of $\ell$.

\begin{figure*}[t]
\begin{center}
\includegraphics[width=0.45\textwidth,clip=true]{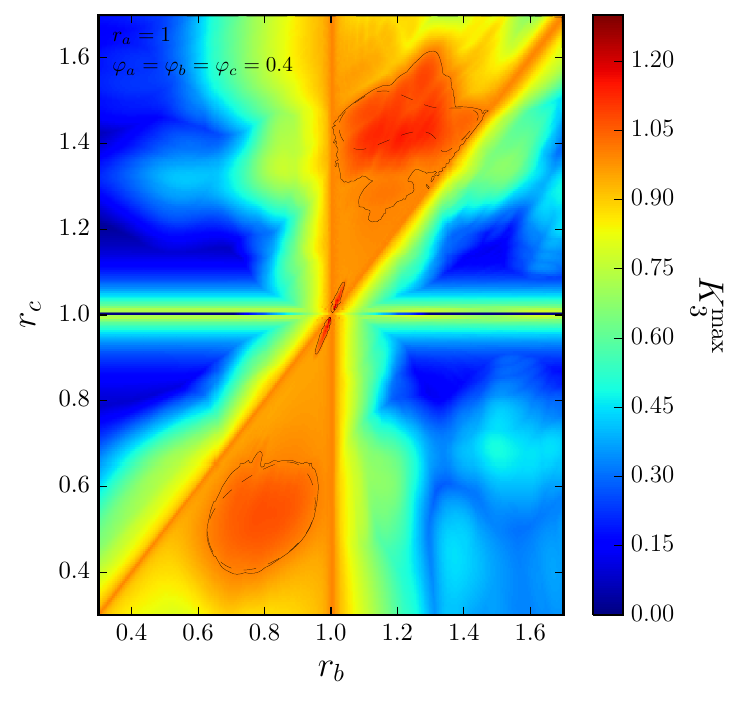}
\includegraphics[width=0.45\textwidth,clip=true]{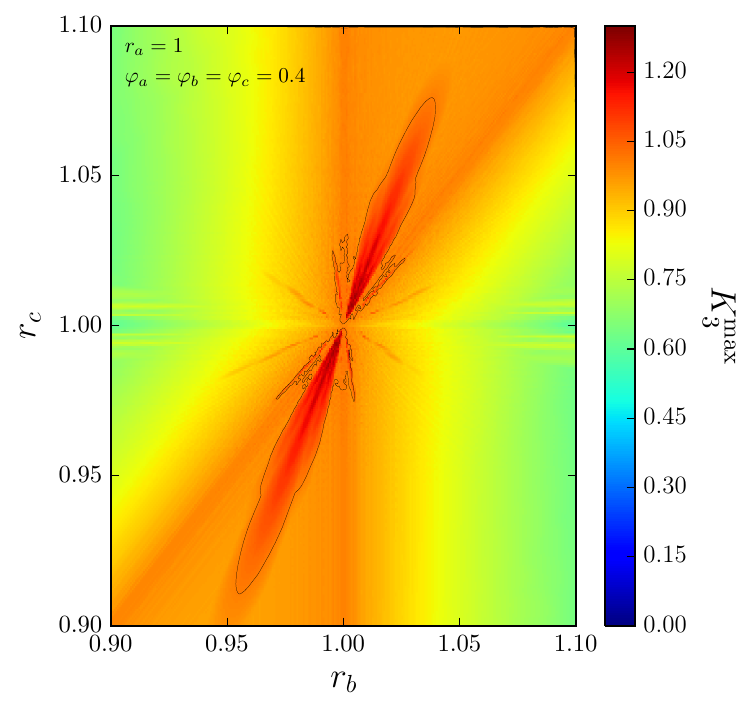}
\includegraphics[width=0.45\textwidth,clip=true]{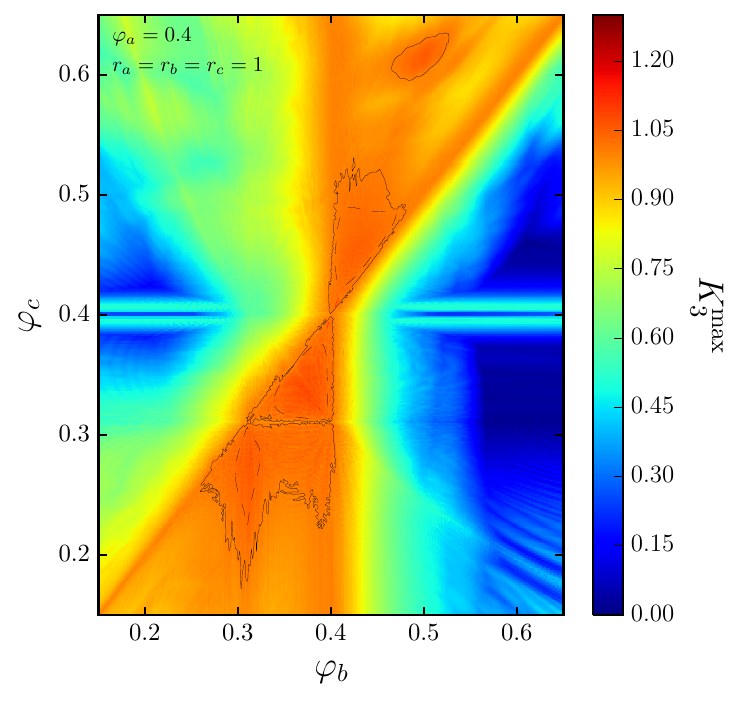}
\includegraphics[width=0.45\textwidth,clip=true]{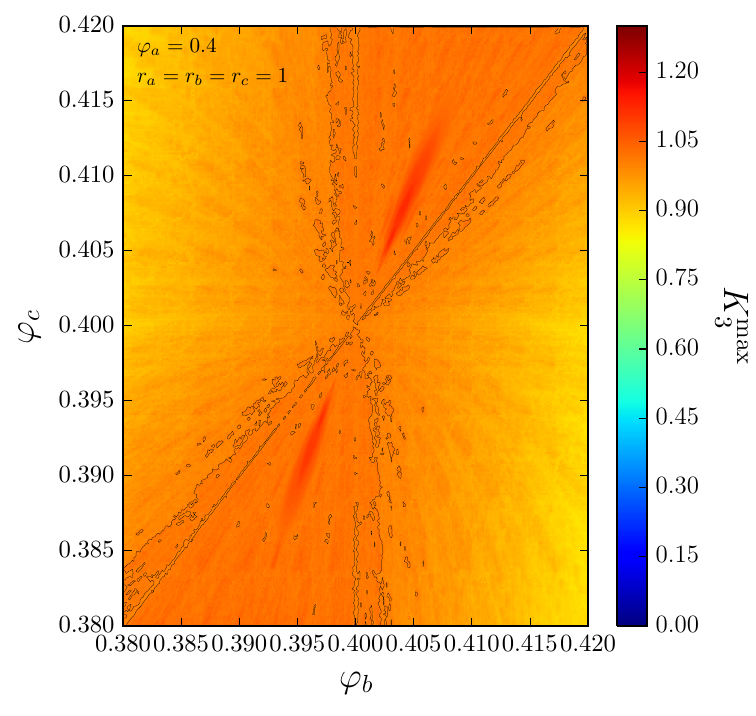}
\includegraphics[width=0.45\textwidth,clip=true]{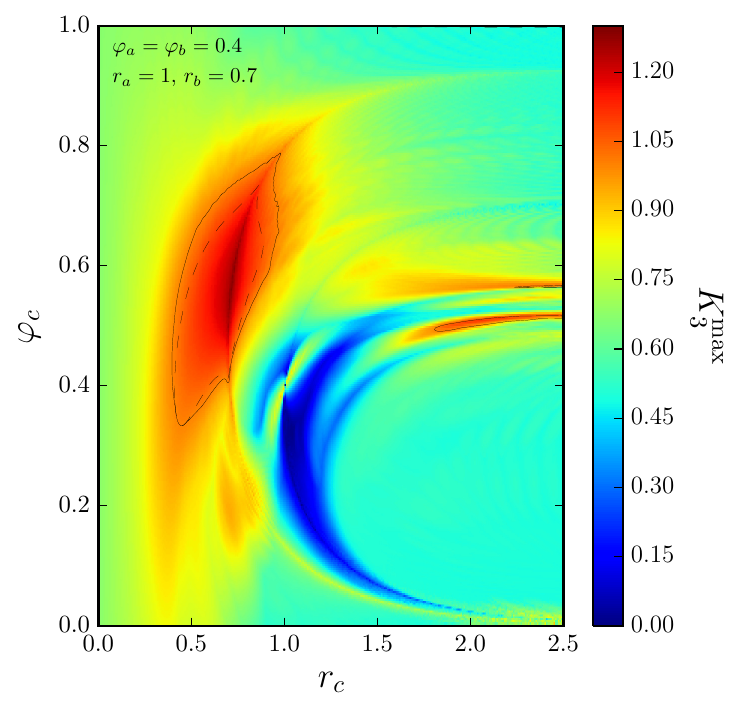}
\includegraphics[width=0.45\textwidth,clip=true]{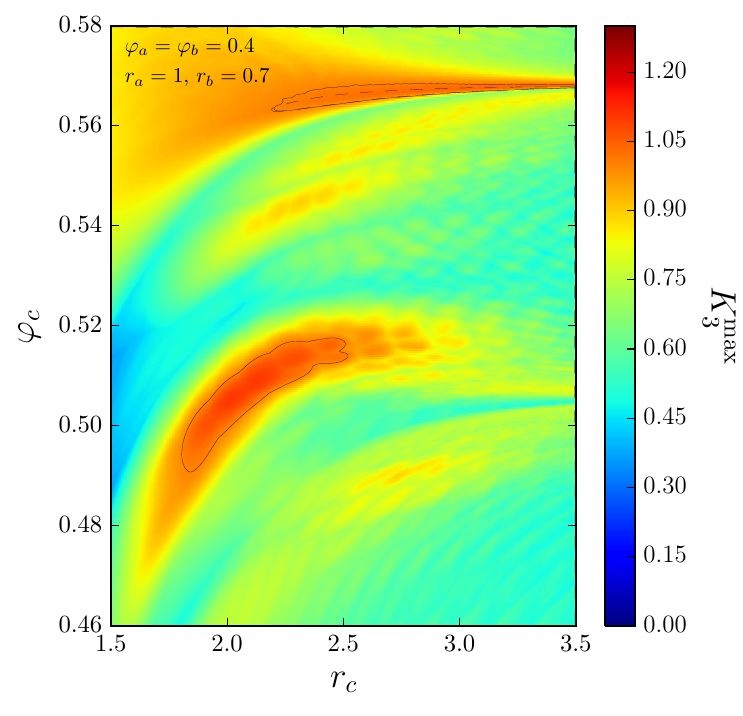}
\caption{Maximal values over $\ell$ of the Leggett-Garg strings $K_3$
  as a function of the squeezing parameters. The black solid lines
  correspond to the contours where the strings equal one and inside
  which LGI violation occurs. The dashed lines stand for the same
  contours but where $\ell$ is taken to infinity instead of maximized
  over. The right panels zoom in on regions of interest of the left
  panels. The top panels show $r_a=1$ and $\varphi_a=\varphi_b=\varphi_c=0.4$, the middle panels 
  $\varphi_a=0.4$ and $r_a=r_b=r_c=1$, and the bottom panels $\varphi_a=\varphi_b=0.4$, $r_a=1$, and $r_b=0.7$.}
\label{fig:maps:1}
\end{center}
\end{figure*}
\begin{figure*}[t]
\begin{center}
\includegraphics[width=0.45\textwidth,clip=true]{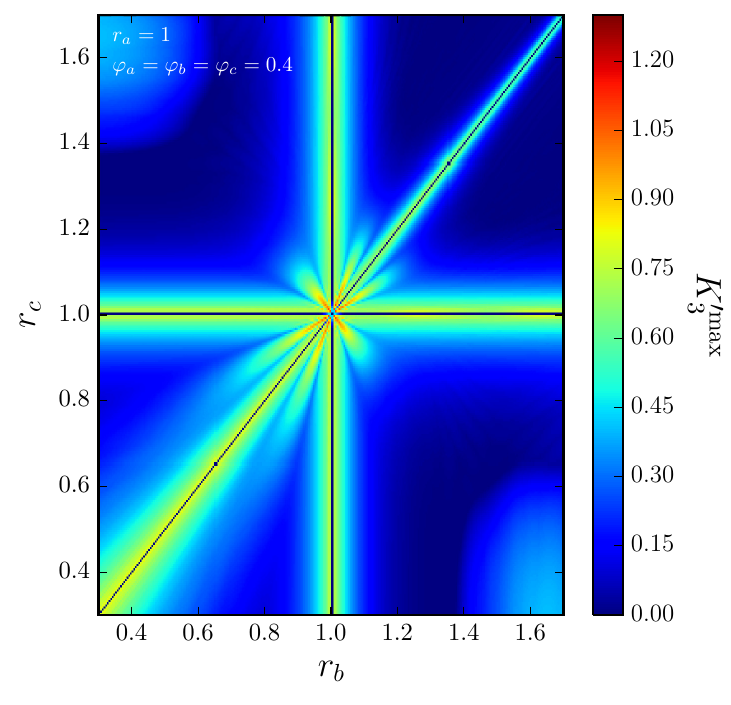}
\includegraphics[width=0.45\textwidth,clip=true]{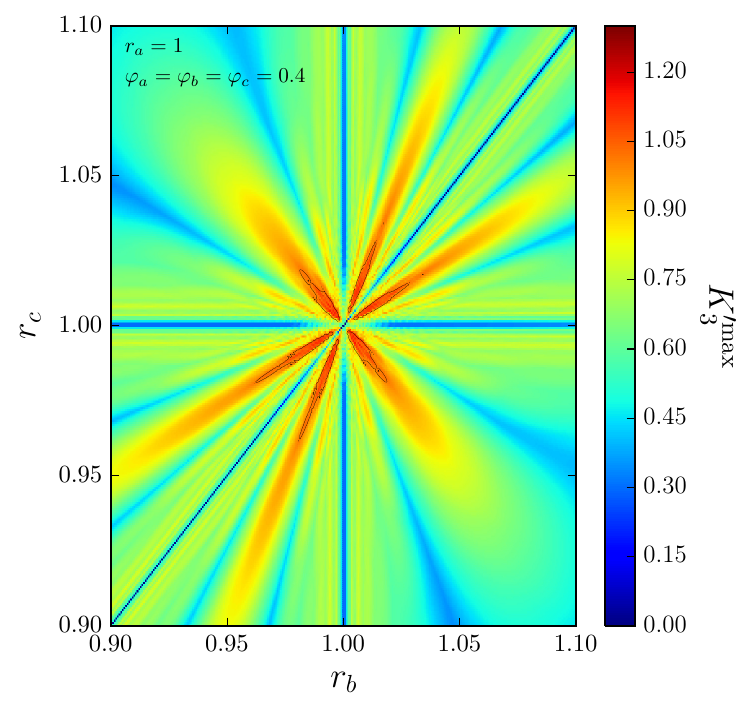}
\includegraphics[width=0.45\textwidth,clip=true]{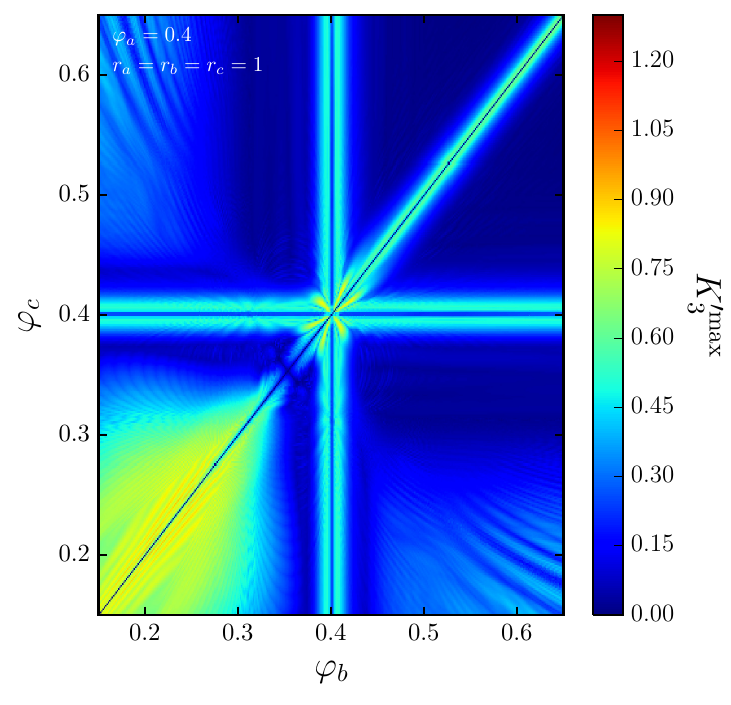}
\includegraphics[width=0.45\textwidth,clip=true]{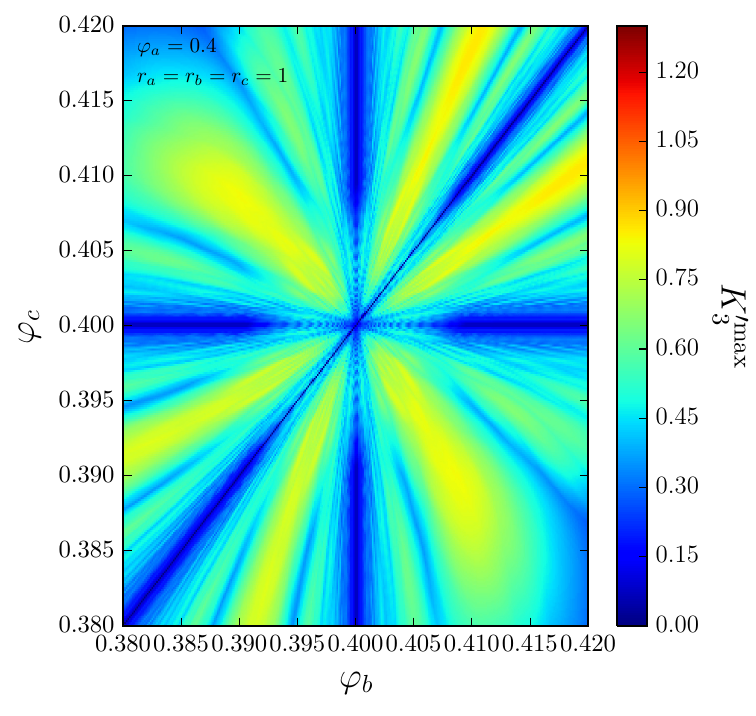}
\includegraphics[width=0.45\textwidth,clip=true]{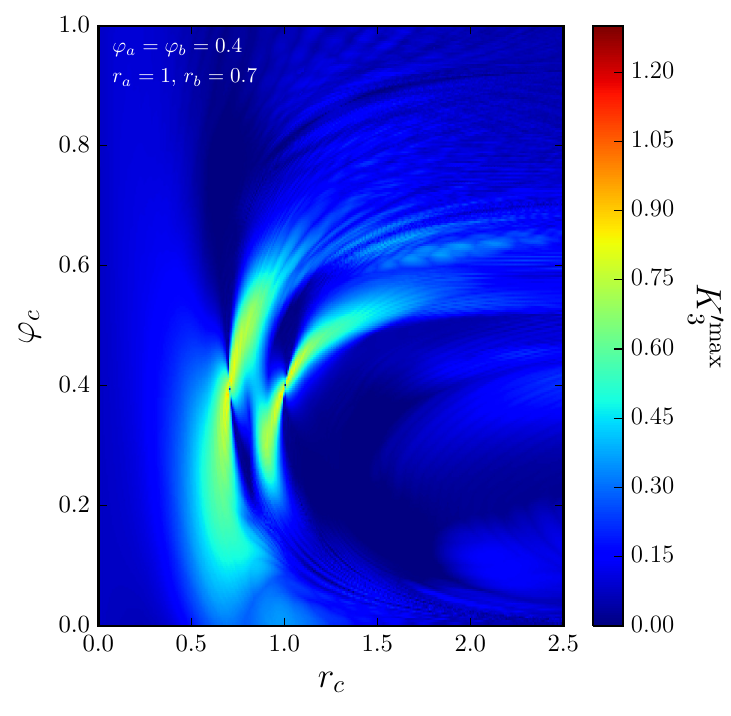}
\includegraphics[width=0.45\textwidth,clip=true]{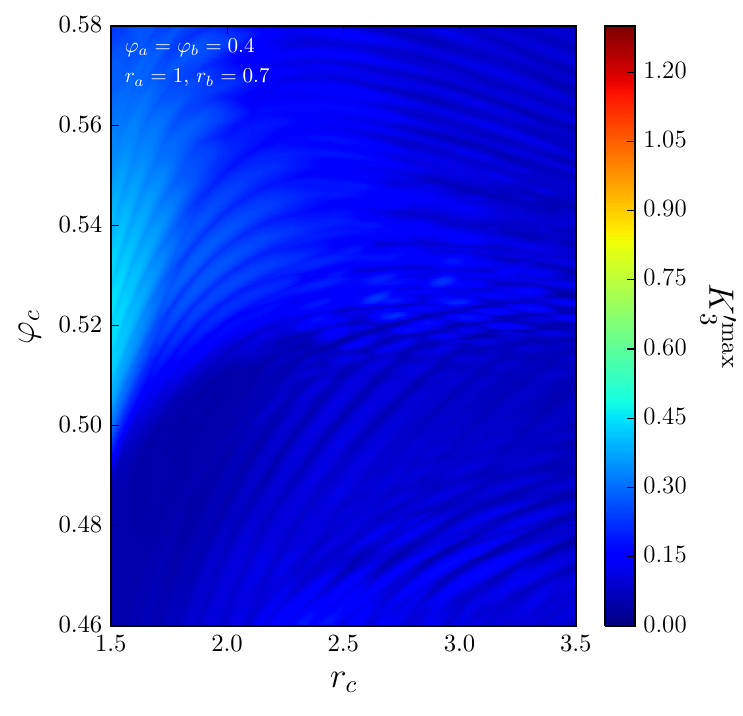}
\caption{Maximal values over $\ell$ of the Leggett-Garg strings $K_3^\prime$
  as a function of the squeezing parameters. The black solid lines
  correspond to the contours where the strings equal one and inside
  which LGI violation occurs. The dashed lines stand for the same
  contours but where $\ell$ is taken to infinity instead of maximized
  over. The right panels zoom in on regions of interest of the left
  panels. The top panels show $r_a=1$ and $\varphi_a=\varphi_b=\varphi_c=0.4$, the middle panels $\varphi_a=0.4$ and
  $r_a=r_b=r_c=1$, and the bottom panels $\varphi_a=\varphi_b=0.4$, $r_a=1$, $r_b=0.7$.}
\label{fig:maps:2}
\end{center}
\end{figure*}
\afterpage{\FloatBarrier}
\clearpage

In \Fig{fig:correlation_r}, we have represented the evolution of the
correlation function versus the squeezing parameter $r$ and the
squeezing angle $\varphi$ for different choices of $\ell$. When
$r_a=r_b$ and $\varphi_a=\varphi_b$, the correlation function is of
course one. Otherwise one can check that it is always between $\pm 1$
and that it tends to zero when $\vert r_b-r_a\vert$ or
$\vert\varphi_a-\varphi_b\vert$ becomes sufficiently large.

\section{Leggett-Garg inequalities}
\label{sec:lgi}
Let us now introduce the so-called $3$-measurement Leggett-Garg
strings $K_3$ and $K_3'$ (which, if needed, could easily be
generalized to $n$-strings) defined by
\begin{align}
\label{eq:K3:def}
K_3 &=C_{ab}+C_{bc}-C_{ac}, \quad K_3'=-C_{ab}-C_{bc}-C_{ac},
\end{align}
where $a$, $b$ and $c$ denote the three times (in chronological
  order) when the measurement is performed. Classical probability
calculus implies $-3\leq K_3, K_3^\prime \leq 1$ and, therefore, any
deviation from those inequalities will be referred to as a LGI
violation. If the state of the system is a qubit $\hat{\sigma}_z$ with
density matrix $\rho=[1+{\bm r}(t)\cdot {\bm \sigma}]/2$ (${\bm r}$ is
the unit Bloch vector and ${\bm \sigma}$ are the Pauli matrices) and
Hamiltonian $\hat{H}=\omega \hat{\sigma}_x/2$ ($\omega$ is the
fundamental frequency of the system), then the correlation function of
$\hat{\sigma}_z$ is simply given by
$C_{ij}=\cos\left[\omega\left(t_i-t_j\right)\right]$. Choosing equal
time intervals $t_c-t_b=t_b-t_a\equiv \tau$, one obtains
$K_3=2\cos(\omega \tau)-\cos(2\omega\tau)$ which clearly violates the
LGI, the maximum violation being obtained for $\omega \tau=\pi/3$ for
which $K_3=3/2$. An important remark is that the correlators $C_{ij}$,
hence the strings $K_3$ and $K_3'$, do not depend on the state
$\bm{r}(t)$ of the qubit. As can be seen in
\Fig{fig:correlation_r}, this is not the case for the squeezed state
where there is a dependence on the parameters characterizing the
  quantum state.

The 3-strings $K_3$ and $K_3^\prime$ are displayed in \Fig{fig:k3} as
a function of $\ell$ for different configurations. One can see that in
some cases (top panel), LGI are not violated, while for others (bottom
panel), there exist values of $\ell$ for which they are. In practice,
making use of the formulas~(\ref{eq:cn1}) and~(\ref{eq:cn2}) for
$C_n$, one can check that when all squeezing angles vanish, no
violation can be obtained. This is in contrast to Bell inequalities
constructed from the same spin operators~\cite{Martin:2016tbd}, where
violation requires $\varphi<0.34\ee^{-r}$ and is maximal precisely for
vanishing squeezing angles. Another important difference between these
two inequalities is that while Bell inequalities violation requires
$r>1.12$~\cite{Martin:2016tbd}, LGI violation occurs even for small
squeezing parameters.

In order to further explore LGI violation in squeezing parameter
space, in~\Figs{fig:maps:1} and~\ref{fig:maps:2}, the maximal values of
$K_3$ and $K_3^\prime$ are displayed as a function of the squeezing
parameters, where maximization is performed over $\ell$. We did not
find configurations for which the classical conditions $K_3,K_3^\prime
\geq -3$ are violated which is why only the maximal values of $K_3$
and $K_3^\prime$ are shown. The right panels zoom in on interesting
features of the left panels. The black solid lines correspond to the
contours $K_3,K_3^\prime=1$, and violation occurs inside them. The
overall structure of these maps is rather complex and usually features
several disconnected regions of parameter space where violation
occurs. Such regions typically correspond to where the different squeezing parameters 
are close but not strictly equal (see the top right and middle right panels of
\Fig{fig:maps:1} and the top right panel of \Fig{fig:maps:2}) but can also exist 
away from these conditions (left panels and bottom right panel in \Fig{fig:maps:1}).
Notice that the non-smooth shapes of the contours
are not numerical artifacts but correspond to genuine irregular
patterns. The dashed lines stand for the contours $K_3,K_3^\prime=1$
but when $\ell$ is taken to infinity instead of maximized over. 
In this case, $K_3$ and $K_3^\prime$ can easily be calculated using \Eq{eq:Cab:ell_eq_infinity}, and as mentioned above, the measurement of $\hat{S}_z$ is simply performed by measuring the sign of the position variable. This regime is therefore experimentally convenient. However,
one can see that in some regions (\ie inside the solid contours but outside
the dashed contours), violation does not occur on the asymptotic
plateau $\ell\rightarrow\infty$ but can be obtained for a bounded
interval of $\ell$ values only. Even though this interval may be
fine-tuned, $\ell$ can be freely chosen by the experimenter so this
does not hamper the practical detection of LGI violation.

More generally, \Figs{fig:maps:1} and~\ref{fig:maps:2} can be used
  to identify the values of $r$ and $\varphi$
  where LGI violations occur. Given a particular experimental
  setting, corresponding to a particular range for $r$ and $\varphi$,
  one can indeed immediately check whether a LGI violation is possible
  or not. As a consequence, these maps hopefully constitute a useful
  guide for designing new experimental protocols.
  
%
\section{Conclusions}
\label{sec:conclusion}
\begin{figure}[t]
\begin{center}
\includegraphics[width=0.35\textwidth,clip=true]{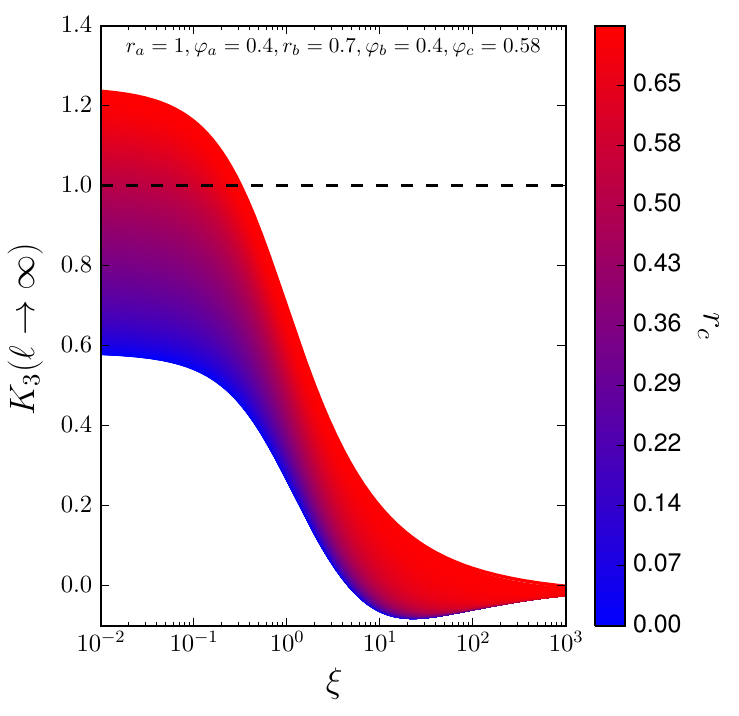}
\caption{Leggett-Garg $3$-string $K_3$ as a function of the decoherence
  parameter $\xi$, in the limit $\ell\rightarrow\infty$ and for a set
  of squeezing parameters, where $r_c$ is varied according to the
  color code. When $\xi$ is of order one or larger, no violation
  occurs.}
\label{fig:decoherence}
\end{center}
\end{figure}
In this paper, we have shown how LGI violation can be obtained with
continuous-variable systems, and have illustrated our approach on the
generic case of squeezed states. By doing so, we have widely extended
the range of systems for which LGI violation can be realized, so far
limited to qubits.

In practice, measuring the spin operator~(\ref{eq:defsz}) from a
measurement of the ``position'' variable $Q$ is straightforward; one
simply needs to determine which integer number $n$ is such that $n\ell
\leq Q < (n+1)\ell$, and take $S_z=(-1)^n$. This is why $\ell$ can be
easily varied over until the largest violation is found and this
parameter can be optimized as done in \Figs{fig:maps:1}
and~\ref{fig:maps:2}. Let us note that if one also has access to
linear combinations of $Q$ and its conjugated momentum $P$, there is
an overall shift in the squeezing angles that one can also optimize
over. Indeed, if one performs a rotation in phase space with angle
$\alpha$ and introduces $Q^\prime=\cos\alpha Q-\sin\alpha P$ and
$P^\prime=\cos\alpha P + \sin\alpha Q$, then the squeezing angles
change according to $\varphi^\prime = \varphi +
\alpha$~\cite{Martin:2016tbd}. As a consequence, if one defines the
pseudo-spin operators with respect to $Q^\prime$ instead of $Q$, one
obtains the same results as the ones derived above but where all the
squeezing angles are shifted by $\alpha$.

Since LGI probe correlations of a single system measured at different
times, quantum decoherence~\cite{Zurek:1981xq, Zurek:1982ii,
  Joos:1984uk} also plays a crucial role for a realistic and practical
experiment~\cite{2010NatPh...6..442P, 2011NatSR...1E.101X}. The effect
of decoherence can be modeled using the quantum channels
formalism~\cite{1999quant.ph.11079K}. For a qubit system, only a few
channels exist and they can be studied
separately~\cite{2013PhRvA..87c2106E, 2015AnPhy.355..241R}. For a
continuous-variable system, however, the dimension of the Hilbert space
is infinite and such a systematic approach cannot be employed without
specifying the environment. In order to assess the impact of
decoherence on our results in a more model-independent way, one can
consider the simple channel in which the density matrix $\rho$ is
mapped onto~\cite{Joos:1984uk}
$\rho(\tilde{Q},\bar{Q})\rightarrow\rho(\tilde{Q},\bar{Q})\exp[-\xi(\tilde{Q}-\bar{Q})^2/2]$,
where the phenomenological parameter $\xi$ encodes the details of the
interaction strength with the environment. In \Eq{eq:Cab:Gauss}, this
amounts to changing $A$ and $B$ according to $A\rightarrow A-\xi/2$
and $B\rightarrow B-\xi$. This models the situation where dynamical
backreaction is small and decoherence is slower than the unitary
evolution of the state. In \Fig{fig:decoherence}, the $3$-string $K_3$
is displayed for a set of squeezing parameters (where $r_c$ is varied
according to the color code) as a function of $\xi$ and in the limit
where $\ell\rightarrow\infty$. When $\xi$ increases, coherence is lost
and $K_3$ is driven to $0$. In the case where LGI are violated at
$\xi=0$, one can see that no violation occurs when $\xi$ is of order
one or larger. This is why limiting the coupling with the environment
is important for a practical implementation of the proposal made in
this paper.

\vspace{0.4cm}
\begin{center}
\textbf{Erratum}\\
$ $\\
\end{center}

After this article was published, we noticed that \Eq{eq:A:def}, in which $A(a,b)$ is defined, contained a typo, which has been corrected in the present version. Since all other formulas do not depend on the precise definition of $A(a,b)$, the rest of the calculation is not affected. However, the figures would need to be redone, since they were derived using the previous version of  \Eq{eq:A:def}. We have nonetheless checked that this does not alter our main conclusion, namely that one can still find configurations where the Leggett-Garg inequalities are violated. For instance, for $r_a = 0.49$, $r_b = 0.77$, $r_c = 1.12$, $\varphi_a = 0.26$, $\varphi_b = 0.27$ and $\varphi_c = 0.4$, the corrected formula gives $K_3\simeq 1.101$ for $\ell=3$.

Moreover, we have realised that, following common practice, we discarded an overall phase in the wavefunction. While measurements performed at the same time are insensitive to such an overall phase, this is not the case for multiple-time measurements
for which the result is sensitive to the change
in the overall phase between the measurement times. See \Refc{Ando:2020kdz} for further details and for a calculation that includes this possible phase change. Our result therefore applies only to the case where the overall phase does not vary in time. Yet, this again does not alter our main conclusion, namely that one can find configurations where the Leggett-Garg inequalities are violated by two-mode squeezed states.
\vspace{0.4cm}
\begin{acknowledgments}
  V.V. acknowledges financial support from STFC grants ST/K00090X/1
  and ST/N000668/1. Numerical computations were done on the Sciama
  High Performance Compute cluster which is supported by the
  ICG, SEPNet and the University of Portsmouth. We thank Kenta Ando for pointing out a typo in \Eq{eq:A:def}. 
\end{acknowledgments}

\appendix
\onecolumngrid

\section{Calculation of the two-point correlation function}
\label{sec:2pt}

In this appendix, we explain how the
  two-point correlation function is calculated. Using the
  expression of the correlator~(\ref{eq:defcorrel}) and the
  definition of the spin operator~(\ref{eq:defsz}), one obtains
\begin{equation}
\label{eq:cab}
C_{ab}=\sum_{n=-\infty}^{n=+\infty}\sum_{m=-\infty}^{m=+\infty}
(-1)^{n+m}\int _{n\ell}^{(n+1)\ell}
\int _{m\ell}^{(m+1)\ell}{\rm d}\tilde{Q}{\rm d}\overline{Q}
\Rea\left[
\Psi_{\rm 1sq}^*(t_a,\tilde{Q})
\Psi_{\rm 1sq}(t_b,\overline{Q})
\langle \tilde{Q}\vert \hat{U}(t_a)\hat{U}^{\dagger}(t_b)
\vert  \overline{Q}\rangle\right] ,
\end{equation}
where $\Psi_{\rm 1sq}$ denotes the (single-mode) squeezed state wave
function given by the following expression
\begin{equation}
\label{eq:wf}
\Psi_{\rm 1sq}(t,Q)=\frac{1}{\pi^{1/4}}\frac{1}{\sqrt{\cosh r}}
\frac{1}{\sqrt{1-z}}
e^{-(1+z)/(1-z)Q^2/2},
\end{equation}
with $z\equiv e^{2i\varphi}\tanh r $. Notice that this wave function
is correctly normalized. Inserting three times the closure relation
for coherent states, the matrix element appearing in
\Eq{eq:cab} can be re-expressed as
\begin{equation}
\langle \tilde{Q}\vert \hat{U}(t_a)\hat{U}^{\dagger}(t_b)
\vert  \overline{Q}\rangle 
=
\int \frac{{\rm d}u}{\pi}
\int \frac{{\rm d}v}{\pi}
\int \frac{{\rm d}w}{\pi}
\langle \tilde{Q}\vert w \rangle
\langle w\vert \hat{U}(t_a)
\vert u \rangle \langle u \vert
\hat{U}^{\dagger}(t_b)
\vert v \rangle \langle v \vert Q \rangle .
\end{equation}
Matrix elements of the form $\langle w\vert \hat{U}(t) \vert u \rangle
$ can be easily calculated using the operator ordering theorem applied to
the evolution operator $\hat{U}$ and quantities such as $\langle v
\vert Q \rangle$ are nothing but the coherent states wave function in
the configuration representation. As a consequence, one arrives at the
following expression
\begin{equation}
\label{eq:matrixUU}
\langle \tilde{Q}\vert \hat{U}(t_a)\hat{U}^{\dagger}(t_b)
\vert  \overline{Q}\rangle 
=\frac{1}{\pi^{7/2}}
\frac{1}{\sqrt{\cosh r_a \cosh r_b}}
e^{-\tilde{Q}^2/2-\overline{Q}^2/2}\int {\rm d}^6\alpha\, 
e^{-\alpha ^{\rm T}M\alpha/2-J^{\rm T}\alpha},
\end{equation}
with $\alpha ^{\rm T}\equiv
[\Rea(u),\Ima(u),\Rea(v),\Ima(v),\Rea(w),\Ima(w)]$ and $J^{\rm
  T}=-\sqrt{2}(0,0,\overline{Q},-i\overline{Q},\tilde{Q},i\tilde{Q})$. The
quantity $M$ is a $6\times 6$ symmetric matrix whose elements can be
written as
\begin{align}
M_{11} &=2-e^{-2i\varphi_a}\tanh r_a-e^{2i\varphi_b}\tanh r_b ,\quad 
M_{12} = -ie^{-2i\varphi_a}\tanh r_a+ie^{2i\varphi_b}\tanh r_b,  \\
M_{13} &=-\frac{1}{\cosh r_b}, \quad M_{14}=-\frac{i}{\cosh r_b},\quad 
M_{15}=-\frac{1}{\cosh r_a}, \quad M_{16}=\frac{i}{\cosh r_a}, \\
M_{22} &= 2+e^{-2i\varphi_a}\tanh r_a+e^{2i\varphi_b}\tanh r_b, \quad   
M_{23} =\frac{i}{\cosh r_b}, \quad M_{24}=-\frac{1}{\cosh r_b},\\
M_{25} &=-\frac{i}{\cosh r_a}, \quad M_{26}=-\frac{1}{\cosh r_a}, \quad
M_{33} =3 +e^{-2i\varphi_b}\tanh r_b, \\
M_{34} &=-i+ie^{-2i\varphi_b}\tanh r_b, \quad M_{35}=M_{36}=0, \quad
M_{44} = 1-e^{-2i\varphi_b}\tanh r_b, \quad M_{45}=M_{46}=0 \\
M_{55} &= 3+e^{2i\varphi_a}\tanh r_a, \quad M_{56}=i-ie^{2i\varphi_a}\tanh r_a, 
\quad M_{66} = 1-e^{2i\varphi_a}\tanh r_a .
\end{align}
From these formula it is straightforward to calculate the determinant
of the matrix $M$. It reads
\begin{equation}
\det M = -128 i \left[ \sin \left(2\varphi_a\right) \tanh r_a 
-\sin \left(2\varphi_b\right) \tanh r_b
-\sin \left(2\varphi_a-2\varphi_b\right)\tanh r_a \tanh r_b\right]\, .
\end{equation}
This determinant vanishes when the two times at which the correlation
function is calculated are the same. Moreover, if
$\varphi_a=\varphi_b=0$ (but, possibly, $r_a\neq r_b$), the
determinant is also zero. These two cases must be treated separately.

Let us first assume that $\det M\neq 0$. Then, the Gaussian
integral~(\ref{eq:matrixUU}) can easily be performed and one finds
\begin{equation}
\label{eq:UU}
\langle \tilde{Q}\vert \hat{U}(t_a)\hat{U}^{\dagger}(t_b)
\vert  \overline{Q}\rangle 
=
\frac{8}{\sqrt{\pi}}
\frac{1}{\sqrt{\cosh r_a \cosh r_b}}
e^{-\tilde{Q}^2/2-\overline{Q}^2/2}
\frac{1}{\sqrt{\det M}}
e^{J^{\rm T}M^{-1}J/2}.
\end{equation}
Clearly, this matrix element is a Gaussian function in $\tilde{Q}$ and
$\overline{Q}$ since $J^{\rm T}M^{-1}J$ is a quadratic form in
$\tilde{Q}$ and $\overline{Q}$, explicitly
\begin{align}
\label{eq:JMJ}
\frac12 J^{\rm T}M^{-1}J &= -\frac{64}{\det M}\left(1-e^{-2i\varphi_a}
\tanh r_a+e^{-2 i \varphi_b}\tanh r_b-e^{2i\varphi_a-2i\varphi_b}
\tanh r_a \tanh r_b \right)\overline{Q}^2 
\nonumber \\ & 
-\frac{64}{\det M}\left(1+e^{2i\varphi_a}
\tanh r_a-e^{2 i \varphi_b}\tanh r_b-e^{2i\varphi_a-2i\varphi_b}
\tanh r_a \tanh r_b \right)\tilde{Q}^2 
+\frac{128}{\det M}\frac{\overline{Q}\tilde{Q}}{\cosh r_a \cosh r_b}.
\end{align}
The final step consists in inserting the above result~(\ref{eq:UU})
into the expression~(\ref{eq:cab}) of the correlation function. This
leads to \Eqs{eq:Cab:Gauss}-(\ref{eq:B:def}). The calculation of the two-point correlation function then reduces to
a double series of terms that are given by the integral of a Gaussian
function over a rectangular domain, the size of which is given by
$\ell$. This series has been computed numerically in order to obtain
the figures of the paper.

Let us now treat the case where $\det M=0$. We first consider the
situation where $t_b\rightarrow t_a$ (meaning $r_b\rightarrow r_a$ and
$\varphi_b \rightarrow \varphi_a$).  In this limit, one can write
\begin{equation}
\label{eq:UUdagger}
\lim _{t_b\rightarrow t_a}\langle \tilde{Q}\vert \hat{U}_a\hat{U}_b^{\dagger}
\vert  \overline{Q}\rangle 
=
\frac{1}{\sqrt{\pi}}
\frac{1}{\cosh r_a \sqrt{\det M}/8}
\exp\left[-\frac{
\left(\overline{Q}-\tilde{Q}\right)^2}
{\left(\cosh r_a \sqrt{\det M}/8\right)^2}\right].
\end{equation}
Then, if we define a small parameter by $\epsilon\equiv \cosh r_a
\sqrt{\det M}/8$ which, obviously, goes to zero since $\det
M\rightarrow 0$, then \Eq{eq:UUdagger} reduces to
\begin{equation}
\lim _{t_b\rightarrow t_a}\langle \tilde{Q}\vert \hat{U}_a\hat{U}_b^{\dagger}
\vert  \overline{Q}\rangle 
=\lim _{\epsilon \rightarrow 0}
\frac{1}{\epsilon \sqrt{\pi}}
e^{-\left(\overline{Q}-\tilde{Q}\right)^2/\epsilon^2}
=\delta \left(\overline{Q}-\tilde{Q}\right).
\end{equation}
As a consequence, in this limit, the correlation
function~(\ref{eq:cab}) takes the form
\begin{align}
\lim _{t_b\rightarrow t_a} C_{ab} & 
=\sum_{n=-\infty}^{n=+\infty}\sum_{m=-\infty}^{m=+\infty}
(-1)^{n+m}\int _{n\ell}^{(n+1)\ell}
\int _{m\ell}^{(m+1)\ell}{\rm d}\tilde{Q}{\rm d}\overline{Q}
\Rea\left[
\Psi_{\rm 1sq}^*(t_a,\tilde{Q})
\Psi_{\rm 1sq}(t_a,\overline{Q})
\delta \left(\overline{Q}-\tilde{Q}\right)\right]\\
& = \sum_{n=-\infty}^{n=+\infty}\int _{n\ell}^{(n+1)\ell}
{\rm d}\tilde{Q}
\Psi_{\rm 1sq}^*(t_a,\tilde{Q})
\Psi_{\rm 1sq}(t_a,\tilde{Q})=1,
\label{eq:cabpsi}
\end{align}
and one verifies that the correlation function is indeed one when the
two times $t_a$ and $t_b$ coincide.

\par

Let us finally focus on the case where $\varphi_a=\varphi_b=\varphi
\rightarrow 0$ but $r_a\neq r_b$.  In this situation, one can define a
new small parameter $\epsilon$ by $\epsilon^2\equiv
-2i\varphi(e^{2r_a}-e^{2 r_b})$ and one has
\begin{equation}
\lim _{\varphi_b\rightarrow \varphi_a}\langle 
\tilde{Q}\vert \hat{U}(t_a)\hat{U}^{\dagger}(t_b)
\vert  \overline{Q}\rangle 
=e^{(r_a+r_b)/2}\lim _{\epsilon \rightarrow 0}
\frac{1}{\epsilon \sqrt{\pi}}
e^{-\left(e^{r_b}\overline{Q}-e^{r_a}\tilde{Q}\right)^2/\epsilon^2}
= e^{(r_a+r_b)/2} \delta \left(e^{r_b}\overline{Q}-e^{r_a}\tilde{Q}\right).
\end{equation}
As a consequence, the two-point correlation function~(\ref{eq:cab})
can now be re-expressed as
\begin{align}
\lim _{\varphi_b \rightarrow \varphi_a} C_{ab} & 
=e^{(r_a+r_b)/2}\sum_{n=-\infty}^{n=+\infty}\sum_{m=-\infty}^{m=+\infty}
(-1)^{n+m}\int _{n\ell}^{(n+1)\ell}
\int _{m\ell}^{(m+1)\ell}{\rm d}\tilde{Q}{\rm d}\overline{Q}
\Rea\left[
\Psi_{\rm 1sq}^*(t_a,\tilde{Q})
\Psi_{\rm 1sq}(t_b,\overline{Q})
\delta \left(e^{r_b}\overline{Q}-e^{r_a}\tilde{Q}\right)\right].
\end{align}
A first integration can be performed thanks to the presence of the
Dirac function. Then the remaining one can also be performed and the
result can be expressed in terms of error functions. This leads to
the formulas given in the main text. It is
interesting to note that the case where the squeezing angles vanish
can only be defined through the limiting procedure explained above
since, taken at face value, the integral in \Eq{eq:matrixUU}
is divergent in this situation.

Finally, let us note that simple expressions can be derived in the
limit $\ell\rightarrow\infty$. In this case, the spin operator
$\hat{S}_z$ defined in \Eq{eq:defsz} is simply the sign operator, \ie
it returns $1$ if $Q \geq 0$ and $-1$ if $Q<0$. In this limit, the
double sum of \Eq{eq:cab} only contains four terms,
corresponding to $(n,m)=(0,0),\,(-1,0),\,(0,-1),\,(-1,-1)$, which are
Gaussian integrals and can therefore be calculated. One obtains \bea
\label{eq:Cab:ellInfinite}
C_{ab}\left(\ell\rightarrow\infty\right) =\Rea\left\lbrace -\frac{4\mathcal{A}(a,b)}{\sqrt{B^2(a,b)-4A(a,b)A^*(b,a)}}\mathrm{arctanh}\left[\frac{B(a,b)}{\sqrt{B^2(a,b)-4A(a,b)A^*(b,a)}}\right]\right\rbrace\, .
\eea

%
\twocolumngrid
\bibliography{lgi}

\end{document}

%% file: newcommands.tex
\newcommand{\ie}{{i.e.~}}

\newcommand{\eg}{{e.g.~}}

\newcommand{\ee}{e}

\newcommand{\Rea}{\Re \mathrm{e}\,}
\newcommand{\Ima}{\Im \mathrm{m}\,}

\newcommand{\beq}{\begin{equation}}
\newcommand{\eeq}{\end{equation}}
\newcommand{\bea}{\begin{eqnarray}}
\newcommand{\eea}{\end{eqnarray}}

\newlength{\wsingfig}
\newlength{\wdblefig}
\newlength{\wquadfig}
\newlength{\wtriplefig}
\newcommand{\Eq}[1]{Eq.~(\ref{#1})}
\newcommand{\Eqs}[1]{Eqs.~(\ref{#1})}
\newcommand{\Fig}[1]{Fig.~{\ref{#1}}}
\newcommand{\Figs}[1]{Figs.~{\ref{#1}}}
\newcommand{\Refc}[1]{Ref.~{\cite{#1}}}
\newcommand{\Refs}[1]{Refs.~{\cite{#1}}}
\newcommand{\Sec}[1]{Sec.~\ref{#1}}